\documentclass{PoS}
\usepackage{subfigure}
\usepackage{xspace}

\newcommand {\APELink} {\textit{APELink}\xspace}
\newcommand {\APENet} {\textit{APENet}\xspace}

\PoS{PoS(LAT2005)100}

\title{Status of the APENet project }

\ShortTitle{Status of the APENet project }

\author{\speaker{Roberto Ammendola}\\
        INFN Roma 2\\
        E-mail: \email{roberto.ammendola@roma2.infn.it}}

\author{Roberto Petronzio\\
        INFN Roma 2 \& Universit\`a di Roma "Tor Vergata"\\
				}

\author{Davide Rossetti\\
        INFN Roma 1\\
        E-mail: \email{davide.rossetti@roma1.infn.it}}
				
\author{Andrea Salamon\\
        INFN Roma 2\\
        E-mail: \email{andrea.salamon@roma2.infn.it}}
				
\author{Nazario Tantalo\\
        INFN Roma 2\\
        E-mail: \email{nazario.tantalo@roma2.infn.it}}
				
\author{Piero Vicini\\
        INFN Roma 1\\
        E-mail: \email{piero.vicini@roma1.infn.it}}
				
\abstract{We present the current status of APENet, our custom
3-dimensional interconnect architecture for PC clusters environment.
We report some micro-benchmarks on our recent large installation as
well as new developments on the software and hardware side.
The low level device driver has been reworked by following a custom
hardware RDMA architecture, and MPICH-VMI, an implementation of the
MPI library, has been ported to APENet.
}

\FullConference{XXIIIrd International Symposium on Lattice Field Theory\\
		 25-30 July 2005\\
		 Trinity College, Dublin, Ireland}

\begin{document}

\section{Introduction}

The \APENet project\cite{APENet,lat04} was started to study the mixing
of existing off-the-shelf computing technology (CPUs, motherboards and
memories for PC clusters) with a custom interconnect architecture,
derived from previous experience of the APE group\footnote{The APE
research group \cite{APE} has traditionally focused on the design and
the development of custom silicon, electronics and software optimized
for Lattice Quantum ChromoDynamics.}. The focus is on building
optimized, super-computer level platforms for LQCD.

%% Traditionally it has been shown that both high bandwidth and low
%% latency are key factors for interconnect performances. Moreover, a
%% multidimensional topology allows a natural decomposition of the
%% lattice, and guarantees a better scaling with respect to an all-to-all
%% topology usually found in commercial interconnects. 

APENet is a three dimensional network of point-to-point links with
toroidal boundary condition. It is characterized by:
\begin{itemize}
\item High bandwidth, over 700MB/s measured on latest Intel Xeon
processors with the stable revision of firmware.
\item Low latency, $\sim 1.9~\mu s$.
\item Natural fit with LQCD and numerical grid-based algorithm; four
dimensional LQCD lattice is easily projected onto the 3D processor
grid.
\item Good performance scaling as a function of the number of
processors; LQCD algorithm mainly use first-neighbor communication so
they scale linearly in the processor count.
\item Very good cost scaling even for large number of processors;
switch-less technology makes the cost function linear in the processor
count.
\end{itemize}

Each computing node is equipped with our custom device, the
\APELink card --- currently at the third hardware version, ---
which is a standard PCI-X 133MHz card with 6 full duplex communication
channels. The main component on the APELink device is a programmable
FPGA, which has many advantages:
\begin{itemize}
\item Low development costs; we avoid the costs --- in the million of
EU range, --- efforts --- two or three experienced engineers ----
and time delay --- one or two years --- typical of custom VLSI
development.
\item It allows easy firmware update on a cabled cluster minimizing
downtime, e.g. to fix bugs.
\item It's possible to add new features and improvements to the
firmware, and install it on already deployed clusters.
\end{itemize}

\begin{figure}[h]
\centering
\includegraphics[width=.7\textwidth]{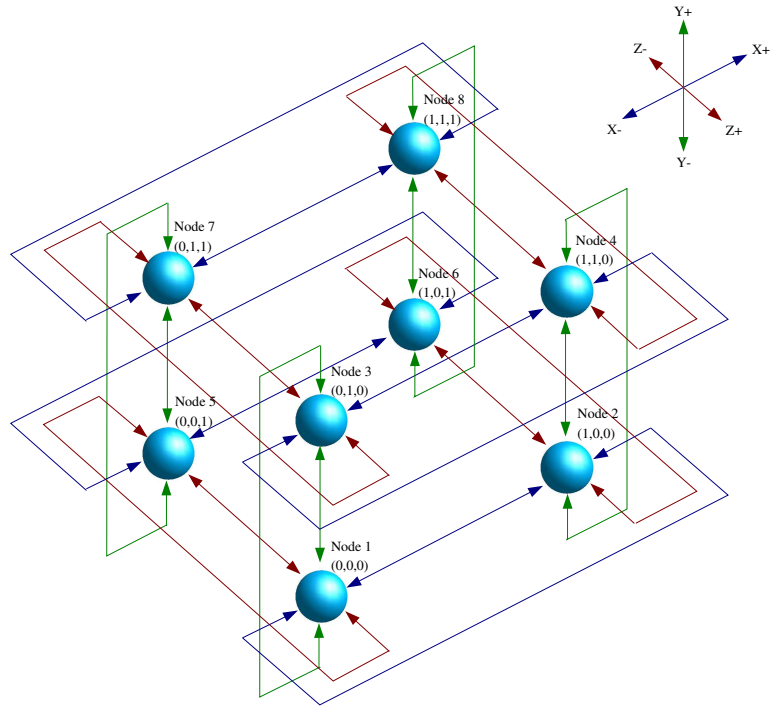}
\caption{A full 3-dimensional torus example, with only two nodes per dimension. All 6 communication channels of each node are connected to other nodes, and the channels can be simultaneously in use. For example if Node 1 and Node 3 are communicating along the Y axis, the flow of packages between Node 7 and Node 4, through Node 3, along the X and Z axis is not affected and can be performed at full speed.}
\label{fig:3d}
\end{figure}

Each \APELink card has internal switching and routing capabilities,
allowing transmission of data packets from one node to any other on the
network --- see figure \ref{fig:3d}. --- The routing mechanism uses a
\textit{dimension ordered} algorithm, which optionally can be replaced
by a \textit{table-driven} user programmable routing. The switching
strategy uses the \textit{wormhole} approach, to achieve minimal
latency in packet handling.

%% \begin{figure}[h]
%% \centering
%% \includegraphics[width=.7\textwidth]{../fig_src/1d}
%% \caption{A unidimensional ring of processing nodes. Data packets flow from a node to an other by hopping through the ring. Hopping is performed in hardware, so that latency is kept to a minimum value.}
%% \label{fig:1d}
%% \end{figure}

%% \begin{figure}[h]
%% \centering
%% \includegraphics[width=.7\textwidth]{../fig_src/2d}
%% \caption{An example of a 2-dimensional torus. Each node is marked with a set of Cartesian coordinates. The path between the source and the destination node is calculated by the minimum distance, and the packets are routed with a dimension ordered algorithm.}
%% \label{fig:2d}
%% \end{figure}

In the following sections we describe the current status of the
project. First we report the latest performance tests on our \APENet
clusters, using the stable version of the firmware and software. Then
we give an overview of the enhancements under development.

\section{Benchmarks}

\begin{figure}
  \centering
  \subfigure[The APE16 cluster]{
    \label{fig:ape16pic}
    \includegraphics[width=.45\textwidth]{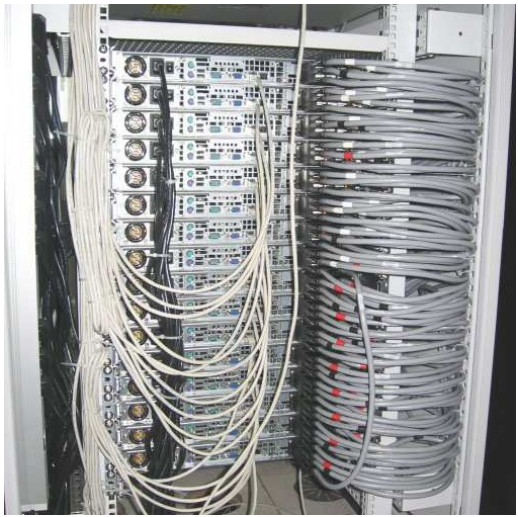}
    %  \caption{APE 16 photo.}
  }
  \hspace{5mm}
  \subfigure[The APE128 cluster]{
    \centering
    \label{fig:ape128}
    \includegraphics[width=.45\textwidth]{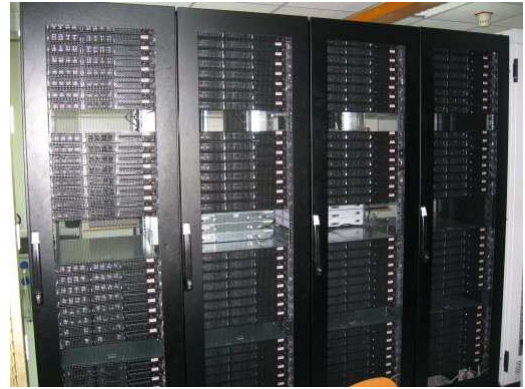}
  }
  \label{fig:ape16draw}
  \caption{The APE128 cluster assembling is still in progress. A total
  of 384 cables will be used, for a total length of more than half
  kilometer.}
\end{figure}

Benchmarks have been done on one testbed (APE16) and on some
processors of a 128 nodes cluster (APE128) which is being deploying as
the time of this writing; both clusters are located in INFN Roma2
computing facility, in the Tor Vergata University:
\begin{description}
\item[APE16] It is a 16 nodes cluster running in Roma2 fully equipped
with a $4 \times 2 \times 2$ APENet topology (rear side is shown in
Fig. \ref{fig:ape16pic}). The processing nodes are dual Xeon 3.0 GHz
with ServerWorks GC-LE chipset and PCI-X at $100 ~MHz$. It runs Fedora
Core 3 in 32bit mode.
\item[APE128] Each processing node is a dual Xeon 3.4 GHz EM64T with
Intel E7320 chipset and $133 ~MHz$ PCI-X bus, running in 64bit mode
under Fedora Core 4 Linux distribution.
\end{description}

Here are presented performance tests on these two setups, based on
standard MPI-level micro-benchmarks \cite{VMI}.

\begin{figure}
\centering
\includegraphics[width=.9\textwidth]{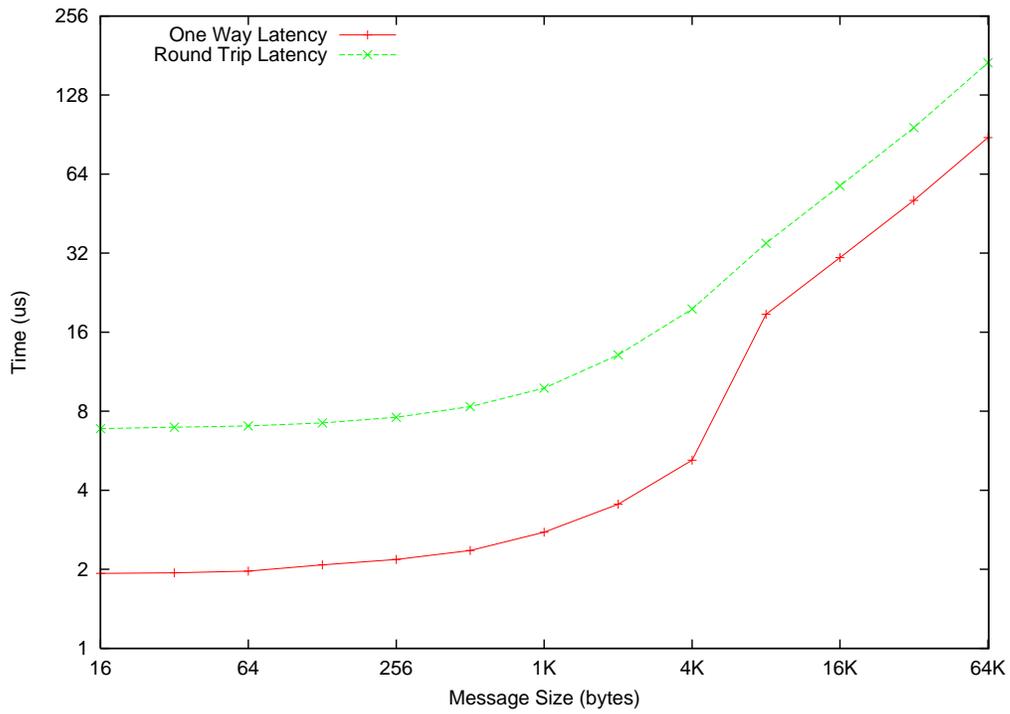}
\caption{MPI latency micro-benchmark: minimum latency for small packets
is $1.9~\mu s$ in the one way case and $6.9~\mu s$ for round trip
time.}
\label{fig:latency}
\end{figure}

For the latency benchmark both the one way and the round trip time are
measured.  In the one way case, all the nodes with even rank perform
an \texttt{MPI\_Send}, while all the nodes with odd rank perform an
\texttt{MPI\_Recv}. Time is taken after $n$ iterations, when a message
is sent back in the opposite direction to synchronize the
processors. This is a streaming test, in which it is stressed the
ability to buffer data and queue commands for multiple subsequent
transmissions.  In the round trip case, the even nodes perform
\texttt{MPI\_Send} $+$ \texttt{MPI\_Recv}, while the odd nodes
\texttt{MPI\_Recv} $+$ \texttt{MPI\_Send}. In this test, the latencies
of the different phases of the transmission process are fully exposed,
while in the previous one they can partly overlap. Time elapse is
averaged after $n$ iterations. Results are plotted in
Fig. \ref{fig:latency}, showing a minimum one way time of $1.9~\mu s$
and a round trip time of $6.9~\mu s$.

\begin{figure}
\centering
\includegraphics[width=.9\textwidth]{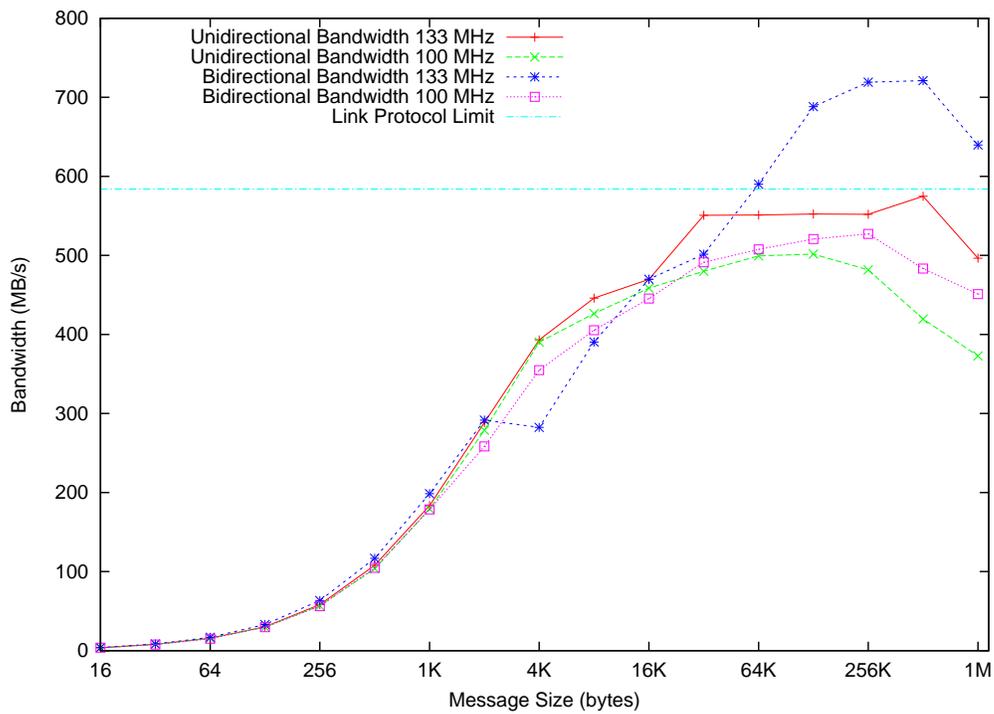}
\caption{MPI bandwidth micro-benchmark: best bandwidth value is over
$700 ~MB/s$ ($1~MB = 1024 \times 1024$ bytes).}
\label{fig:bandwidth}
\end{figure}

Two bandwidth benchmarks have been performed: unidirectional,
\texttt{MPI\_Send} for even nodes and \texttt{MPI\_Recv} for odd
nodes, and bidirectional, where all the nodes perform an
\texttt{MPI\_Sendrecv}. Results are plotted in
Fig.\ref{fig:bandwidth}. For the unidirectional case a peak value of
$\sim~570~MB/s$ have been measured, which represents more than $90 \%$
of the single channel theoretical bandwidth of $585~MB/s$, which is
the limiting factor in this case. The bidirectional case gives a best
value of $\sim~720~MB/s$ for big buffer sizes; here the theoretical
limit is fixed by the PCI-X bus bandwidth, which is $1015~MB/s$.

% \begin{figure}%[htp]
%      \centering
%      \subfigure[Latency]{
%           \label{fig:lat}
%           \includegraphics[width=.45\textwidth]{../data/latency}}
% %     \hspace{.3in}
%      \subfigure[Bandwidth]{
%           \label{fig:band}
%           \includegraphics[width=.45\textwidth]{../data/bandwidth}}\\
% %     \vspace{.3in}
% %     \hspace{.1in}
%      \subfigure[Latency]{
%           \label{fig:lat2}
%           \includegraphics[width=.45\textwidth]{../data/latency}}
%      \subfigure[Latency]{
%           \label{fig:lat3}
%           \includegraphics[width=.45\textwidth]{../data/latency}}
%      \caption{Insieme di figure}
%      \label{fig:multifig}
% \end{figure}

\section{Latest Improvements}
A major rework of the card PCI DMA controller --- the firmware block
responsible for interaction with the PCI bus and main computer memory
--- and consequently of low level device driver has been done. The
main goal of this activity is reducing the need for the CPU to access
the card for packet receiving and transmitting. This way the CPU has
more cycles to be spent on the number-crunching task and the \APELink
card is more independent, especially in the packet receiving process.
\begin{itemize}
\item On the receiver part of the PCI logic, a RDMA (Remote Direct
Memory Access) approach has been developed. In a 64 bit-wide dual port
RAM, the driver stores the addresses of a set of published buffers;
\item When a data packet is sent, it points to a certain buffer ID, so
that the DMA on the receiver side can be performed without involvement
of the local CPU.
\item On the transmitting side, a scatter/gather FIFO is used to
minimize target accesses to the PCI Base Address Registers. This FIFO
can gather the instruction to perform DMAs of various types, allowing
multiple queues.
\item There is hardware support for link multiplexing. Each card
supports the abstraction of the \textit{port} and there exist up to 4
\textit{ports}.
\item A standard MPI layer is now available. A porting of the
MPICH-VMI \cite{VMI} has been developed which fully exploits the RDMA
architecture.
\end{itemize} 

The multiplexing of the \APELink card is especially important to fully
exploit the two CPU available on each motherboard. Typically two
process instances are spawn on each motherboard and they have to share
the \APELink card and have the \APENet traffic properly
dispatched. Furthermore, we plan to reserve one \textit{port} to carry
TCP/IF protocol traffic on it, which is a planned feature to be added.

We are also working on the execution environment which is really
necessary for a large cluster. We are providing cluster partitioning
and integration with standard batch queueing systems (PBS, Torque,
...). The idea is that the 3D grid of processors can be split into
subsets which are still topologically connected, e.g. a
$8\times4\times4$ 3D torus can be split into 8 $4\times4$ independent
partitions having 2D topology.
% Different partitioning, e.g. 2 $4\times4\times4$ independent systems, are possibile only by re-cabling, which is quite time consuming.

%% \begin{figure}
%% \centering
%% \includegraphics[width=.7\textwidth]{../fig_src/archi}
%% \caption{Card structure.}
%% \label{fig:card}
%% \end{figure}

\section{Conclusions}

The latency results can be considered pretty fine compared with actual
commercial interconnects. Even unidirectional bandwidth is quite close
to its theoretical limit in machines with $133 ~MHz$ on the PCI-X
bus. For bidirectional bandwidth, the measured values show that there
is still room for improvement. We believe that the enhancements under
development can give a substantial performances boost, in particular
for smaller message sizes. The APE128 cluster (128 nodes, with
topology $8 \times 4 \times 4$) has been deployed and is being cabled
(pictures in Fig. \ref{fig:ape128}). We anticipate that real scaling
benchmarks of LQCD applications will be performed on it, as long as
competitive physics production.

%% \begin{figure}
%%   \centering
%% %  \subfigure[Scheme of the $8 \times 4 \times 4$ topology cabling. A total of 384 cables will be used, for a total length of more than half kilometer.]{
%%     \label{fig:ape128cabling}
%%     \includegraphics[width=.45\textwidth]{../fig_src/cabling}
%%   \caption{APE 128.}
%%   \label{fig:ape128}
%% \end{figure}

\end{document}